%% file: Datafilter.tex
\documentclass[preprint,12pt]{elsarticle}
\usepackage{epsfig,url}
\usepackage{lineno}
\usepackage{orcidlink}

\raggedright

\begin{document}

\begin{frontmatter}

\title{The Online Data Filter for the KM3NeT Neutrino Telescopes}

\include{Authors}

\include{Abstract}

\end{frontmatter}

\include{Introduction}

\include{Trigger}

\include{Performance}

\include{Conclusions}

\include{Acknowledgements}

\bibliographystyle{elsarticle-num}

\bibliography{Datafilter}
  
\end{document}

%% file: Authors.tex
\newcommand\blfootnote[1]{%
  \begingroup
  \renewcommand\thefootnote{}\footnote{#1}%
  \addtocounter{footnote}{-1}%
  \endgroup
}

\newcounter{daggerfootnote}
\newcommand*{\daggerfootnote}[1]{%
    \setcounter{daggerfootnote}{\value{footnote}}%
    \renewcommand*{\thefootnote}{\fnsymbol{footnote}}%
    \footnote[2]{#1}%
    \setcounter{footnote}{\value{daggerfootnote}}%
    \renewcommand*{\thefootnote}{\arabic{footnote}}%
}



\cortext[cor]{corresponding author}

\author[b,a]{O.~Adriani\,\orcidlink{0000-0002-3592-0654}}
\author[c]{S.~Aiello}
\author[d,bd]{A.~Albert}
\author[e]{A.\,R.~Alhebsi\,\orcidlink{0009-0002-7320-7638}}
\author[f]{M.~Alshamsi}
\author[g]{S. Alves Garre\,\orcidlink{0000-0003-1893-0858}}
\author[i,h]{A. Ambrosone}
\author[j]{F.~Ameli}
\author[k]{M.~Andre}
\author[l]{L.~Aphecetche\,\orcidlink{0000-0001-7662-3878}}
% ...
\address[a]{Universit{\`a} di Firenze, Dipartimento di Fisica e Astronomia, via Sansone 1, Sesto Fiorentino, 50019 Italy}
\address[b]{INFN, Sezione di Firenze, via Sansone 1, Sesto Fiorentino, 50019 Italy}
\address[c]{INFN, Sezione di Catania, (INFN-CT) Via Santa Sofia 64, Catania, 95123 Italy}
\address[d]{Universit{\'e}~de~Strasbourg,~CNRS,~IPHC~UMR~7178,~F-67000~Strasbourg,~France}
\address[e]{Khalifa University of Science and Technology, Department of Physics, PO Box 127788, Abu Dhabi,   United Arab Emirates}
\address[f]{Aix~Marseille~Univ,~CNRS/IN2P3,~CPPM,~Marseille,~France}
\address[g]{IFIC - Instituto de F{\'\i}sica Corpuscular (CSIC - Universitat de Val{\`e}ncia), c/Catedr{\'a}tico Jos{\'e} Beltr{\'a}n, 2, 46980 Paterna, Valencia, Spain}
\address[h]{INFN, Sezione di Napoli, Complesso Universitario di Monte S. Angelo, Via Cintia ed. G, Napoli, 80126 Italy}
\address[i]{Universit{\`a} di Napoli ``Federico II'', Dip. Scienze Fisiche ``E. Pancini'', Complesso Universitario di Monte S. Angelo, Via Cintia ed. G, Napoli, 80126 Italy}
\address[j]{INFN, Sezione di Roma, Piazzale Aldo Moro, 2 - c/o Dipartimento di Fisica, Edificio, G.Marconi, Roma, 00185 Italy}
\address[k]{Universitat Polit{\`e}cnica de Catalunya, Laboratori d'Aplicacions Bioac{\'u}stiques, Centre Tecnol{\`o}gic de Vilanova i la Geltr{\'u}, Avda. Rambla Exposici{\'o}, s/n, Vilanova i la Geltr{\'u}, 08800 Spain}
\address[l]{Subatech, IMT Atlantique, IN2P3-CNRS, Nantes Universit{\'e}, 4 rue Alfred Kastler - La Chantrerie, Nantes, BP 20722 44307 France}
% ...
\address[bd]{Universit{\'e} de Haute Alsace, rue des Fr{\`e}res Lumi{\`e}re, 68093 Mulhouse Cedex, France}
% ...

%% file: Abstract.tex
\begin{abstract}
  The KM3NeT research infrastructure comprises two neutrino telescopes located in the deep waters of the Mediterranean Sea, namely ORCA and ARCA.
  KM3NeT/ORCA is designed for the measurement of neutrino properties and
  KM3NeT/ARCA for the detection of high\nobreakdashes-energy neutrinos from the cosmos.
  Neutrinos are indirectly detected using three\nobreakdashes-dimensional arrays of photo\nobreakdashes-sensors
  which detect the Cherenkov light that is produced when relativistic charged particles emerge from a neutrino interaction.
  The analogue pulses from the photo\nobreakdashes-sensors are digitised offshore and all digital data are sent to a station on shore
  where they are processed in real time using a farm of commodity servers and custom software.
  In this paper, the design and performance of the software that is used to filter the data are presented.
  The performance of the data filter is evaluated in terms of its
  efficiency, 
  purity and
  capacity.
  The efficiency is measured by the effective volumes of the sensor arrays as a function of the energy of the neutrino.
  The purity is measured by a comparison of 
  the event rate caused by muons produced by cosmic ray interactions in the Earth’s atmosphere with
  the event rate caused by the background from decays of radioactive elements in the sea water and bioluminescence.
  The capacity is measured by the minimal number of servers that is needed to sustain the rate of incoming data.
  The results of these evaluations comply with all specifications.
  The count rates of all photo\nobreakdashes-sensors are measured with a sampling frequency of $10~\mathrm{Hz}$.
  These data are input to the simulations of the detector response and will also be made available for interdisciplinary research.
\end{abstract}

%% file: Introduction.tex
\section{Introduction}

The KM3NeT research infrastructure comprises two neutrino telescopes located in the deep waters of the Mediterranean Sea, namely ORCA and ARCA \cite{km3net-LOI}.
They are located off the shores of Toulon, France and Portopalo di Capo Passero (Sicily), Italy, respectively.\\[\baselineskip]
The configuration of ORCA is optimised for the study of atmospheric neutrinos with energies between $2-100~\mathrm{GeV}$ and 
the determination of the neutrino mass ordering.
The configuration of ARCA is optimised for the detection of neutrinos from the cosmos with energies in excess of $1~\mathrm{TeV}$.

Neutrinos are indirectly detected via the Cherenkov light that is produced when relativistic charged particles emerge from a neutrino interaction.
The chosen photo\nobreakdashes-sensor is a 3\nobreakdashes-inch photo\nobreakdashes-multiplier tube (PMT) which combines 
accurate timing,
high quantum efficiency,
low dark count and
large area \cite{km3net-PMT}.
To operate the PMTs in the deep sea,
31 of them are housed in a pressure\nobreakdashes-resistant glass sphere, referred to as an optical module \cite{km3net-DOM}.
Each optical module also houses the necessary electronics to process the signals from the PMTs and to send the data to shore.
To deploy the optical modules,
18 of them are mounted on a narrow flexible structure, referred to as a string \cite{km3net-LOM}.
The strings are anchored to the seabed and kept vertical by the buoyancy of the optical modules and a $1.35~\mathrm{kN}$ buoy on top.
The spacing between the optical modules in the ARCA and ORCA neutrino telescopes is optimised for the energies of interest.
The vertical spacing between the optical modules is about $36~\mathrm{m}$ and $9~\mathrm{m}$ and
the horizontal spacing between the strings about $90~\mathrm{m}$ and $20~\mathrm{m}$, respectively.
A set of strings is operated as a standalone detector.
A total of 115 strings is referred to as a building block.
The geometrical volume of a building block is defined by the smallest cylinder encompassing all PMTs in the detector.
For ARCA and ORCA, it is about
$5.6 \times 10^{-1}~\mathrm{km^{3}}$ and
$7.6 \times 10^{-3}~\mathrm{km^{3}}$, respectively.
The ARCA and ORCA neutrino telescopes are under construction.
At completion, they will consist of two and one building blocks, respectively.
The telescopes are operated during construction.\\[\baselineskip]

The interaction of a neutrino, the passage of a muon or other signals of interest are referred to as events.
The primary task of the online data filter is 
to detect the wanted events and 
to reject the background due to decays of radioactive elements in the sea water and bioluminescence.
An overview of the data processing system is presented in section \ref{Data processing}, covering
the offshore data acquisition,
the data transmission,
the on shore data processing and
the data recording.
The definitions of the various types of data that are recorded are introduced in this section.
The top\nobreakdashes-level specifications are summarised in section \ref{Specifications}.
The event signatures and the main algorithm are presented in section \ref{Data filter}.
The performance of the online data filter is assessed in section \ref{Performance} and
the conclusions are summarised in section \ref{Conclusions}.

\section{Data processing}
\label{Data processing}

\begin{figure}[h]
  \begin{center}
    \begin{picture}(10,10)
      \put(-0.5,0){\includegraphics*[trim={5cm 1cm 6cm 0cm},clip,height=10cm]{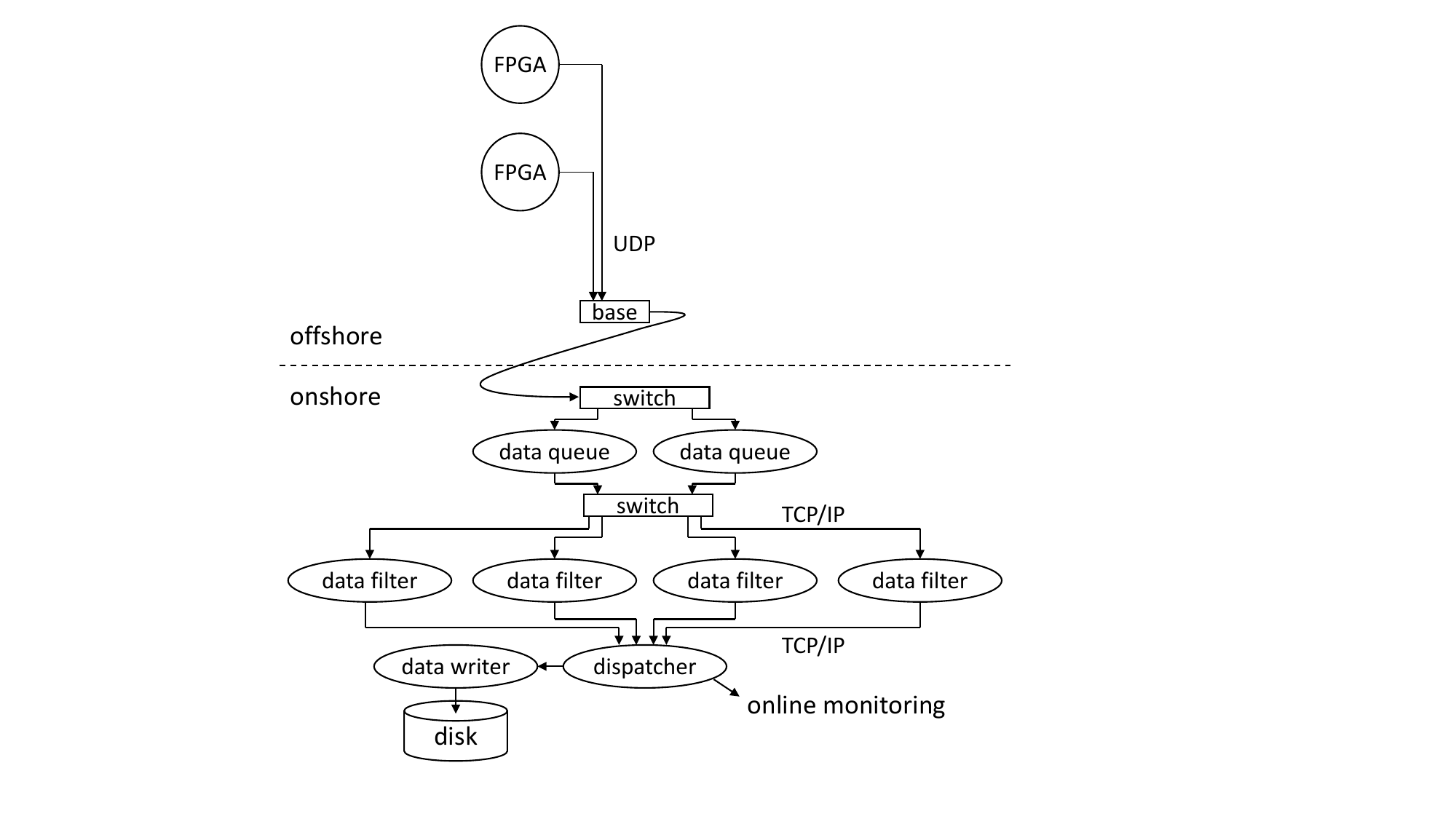}}
    \end{picture}
  \end{center}
  \caption{\label{fig:schematic-view}
    Schematic view of the data flow (not to scale).
    The part above the dashed line corresponds to offshore and the part below to on shore.
    The circles correspond to the optical modules which are deployed in the deep sea and house the PMTs as well as the FPGA.
    The ellipses and rectangular boxes correspond to processes and switch fabric, respectively.
    The base provides for the connection to the sea\nobreakdashes-floor network and is mounted on the anchor of the string.
  }
\end{figure}

The data processing system of KM3NeT is based on the all\nobreakdashes-data\nobreakdashes-to\nobreakdashes-shore concept.
The analogue pulses from the PMTs are digitised offshore and all digital data are sent to shore 
where they are processed in real time using a farm of commodity servers and custom software (see figure \ref{fig:schematic-view}).
This concept has been successfully pioneered in the ANTARES project \cite{ANTARES-DAQ}.

\subsection{Offshore data acquisition}

The analogue output of a PMT is processed via
an amplifier with pulse shaping and
a time-over-threshold discriminator.
The typical threshold of the discriminator corresponds to 0.25 photo-electron equivalent ($\mathrm{p.e.}$).
Each analogue pulse that passes this threshold is digitised.
The digital data contain 
a PMT identifier (1 byte), 
the time stamp of the leading edge (4 bytes) and
the length of the pulse (1 byte),
jointly referred to as a hit.
The least significant bits of the time stamp and the pulse length correspond to $1~\mathrm{ns}$.
The dynamic ranges of the time stamp and pulse length are thus $2^{32}~\mathrm{ns}$ and $2^8~\mathrm{ns}$, respectively.
The digitisation of the signals from the different PMTs is implemented inside a field programmable gate array (FPGA) that is located inside the optical module \cite{JATIS}.
The FPGA also hosts a microprocessor which is connected to a station on shore via a fibre\nobreakdashes-optic network.
The time stamp is obtained from a local clock that runs inside the FPGA which is synchronised to a master clock on shore 
using the White Rabbit system \cite{WR}.
The time stamp from the local clock is reset every $100~\mathrm{ms}$.
The master clock is locked to GPS.
The actual time is maintained in UTC and included in the data throughout the processing.
The count rate of a PMT typically amounts to about $7~\mathrm{kHz}$ \cite{km3net-PPMDU}.
This rate is primarily due to decays of radioactive elements in the sea water but 
can increase for certain periods due to varying contributions from bioluminescence.
The dark count rate {\em in situ} is a few hundred Hertz.
The total data rate amounts to about $25~\mathrm{Gb/s}$ per building block.

\subsection{Data transmission}

For the distribution of the data on shore,
the continuous data streams from the optical modules are sliced in time by the FPGA.
The duration of the time slice is set to $100~\mathrm{ms}$ which matches the reset of the time stamp.
The data are organised into jumbo frames of $9000$ bytes using a FIFO and sent to shore using UDP.
Each slice in time yields about $15$ jumbo frames.
Counters are included in the header of each jumbo frame to reassemble the UDP frames on shore and
to monitor possible losses of UDP packets.
To limit the probability of packet losses somewhere in the network due to anomalous high count rates in some of the PMTs,
the maximal number of hits per PMT that can be recorded during a time slice is set to $2000$.
The remaining data are sent to shore anyway and are used for monitoring.
This limit corresponds to $20~\mathrm{kHz}$ and is referred to as the high\nobreakdashes-rate veto.
A temporary increase of the background due to a burst of bioluminescence is thereby filtered.
As a result,
the probability of a packet loss is normally less than $10^{-6}$.
Seasonal variations of the bioluminescence have been observed at the site of ORCA which lead to a general increase of the background.
This is included in the specifications of the data filter.
Due to the high\nobreakdashes-rate veto,
the total data rate is limited any time to about $75~\mathrm{Gb/s}$ per building block.

\subsection{On shore data processing}

Three software applications have been developed in the C$^{++}$ programming language to process the data on shore, namely
``data queue'',
``data filter'' and
``data writer''.
Multiple instances of the data queue and data filter applications run in parallel on the servers.
The neutrino telescopes as well as the applications to process the data are configured and operated via the control unit \cite{km3net-control-unit}.
All data corresponding to the same time slice are sent to the same data filter.
For each time slice,
the UDP frames that originated from the same optical module are first assembled into a single frame by one of the data queues and
then distributed to the data filters using TCP/IP.
The distribution is based on the UTC time of the data and sustained using round\nobreakdashes-robin scheduling of the data filters.
As a prerequisite,
the time to process a slice of data by a data filter must be shorter than
the duration of the time slice multiplied by the number of data filters.
The farm of servers should thus provide the capacity needed to process the incoming data in real time.
All data produced by the data filters are sent via a dispatcher to one central data writer using TCP/IP.
Normally, the neutrino telescopes are operated around the clock.
Apart from sea operations during the construction of the telescope, maintenance of the shore stations and taking of calibration data,
there is no significant dead\nobreakdashes-time.
During operation, data are continuously recorded using consecutive runs of typically 3 hours, 
each of which is identified by a unique number.
For each run, 
a separate file is written by the data writer which is archived in multiple computer centres.
Each of these files represents a single set of data with a well\nobreakdashes-defined and reproducible configuration of the whole system.\\[\baselineskip]

An event is triggered by the identification of a cluster of hits which complies with predefined criteria.
Following a trigger, all telescope data within a time window around the time of the event are recorded.
The reduction of the data is thus given by the product of duration of this time window and the trigger rate.
Different trigger algorithms are operated simultaneously.
In addition,
level\nobreakdashes-0 (L0),
level\nobreakdashes-1 (L1),
level\nobreakdashes-2 (L2),
Supernova (SN), and
summary data
can be recorded according to predefined samplings.
The L0 data contain all hits from all PMTs in the detector.
The L1 data are a subset of the L0 data and contain only local coincidences between two (or more) L0 hits in one optical module within a predefined time window.
The L2 and SN data are a subset of the L1 data subject to additional constraints.
These constraints apply to the time difference between hits, the space angle between PMT axes and the number of hits.
The L0, L1 and L2 data correspond to the first processing steps of the data filter.
They are used for monitoring of the data quality and for the calibration of the detector.
The SN data are used to study low\nobreakdashes-energy neutrinos ($\mathrm{MeV}$) from e.g.\ Supernovae.
The criteria for L0, L1, L2 and SN data are similar for ORCA and ARCA.
The summary data contain the rate and status of the offshore data acquisition of all PMTs in the detector per slice in time.
They are used to monitor
the operation of the detector and the conditions in the deep sea.
To limit the volume of the summary data,
the measured rate of each PMT is compressed to a single byte
using the logarithmic value of the counts per unit time.
The dynamic range is from $2~\mathrm{kHz}$ ($1$) to $2~\mathrm{MHz}$ ($256$).
A zero value ($0$) corresponds to any rate that is less than the lower limit.
The high\nobreakdashes-rate veto and FIFO status from the FPGA are recorded for each PMT inside a given optical module using two values of 32 bits.
The UDP frame counters are included in the summary data using two values of 16 bits.

\subsection{Data recording}

Normally, all triggered events, all summary data and all SN data are recorded.
In addition, a small fraction of the L1 data (typically 1\%) are recorded for the calibration of the PMTs.
Optionally, each data filter can also maintain a circular buffer containing the L0, L1, L2 or SN data
from a predefined number of the latest time slices.
Following the reception of an external alert (e.g.\ from other telescopes),
these data are written to the local disks of the servers and collected afterwards for offline analyses.
The maximal time that can be recovered depends on the available random access memory (RAM) in the CPUs of the servers and could reach several minutes of history.\\[\baselineskip]
The neutrino telescope data are filtered online using calibrations that have been produced beforehand.
These calibrations are archived in the central database and communicated to the data filters by the control unit.
The calibrations include the position, orientation and time offset of each PMT.
Since the filtered data contain an unmodified copy of the original raw data,
the recorded data can be analysed offline using calibrations that are produced afterwards.
The attainable resolution of the neutrino telescope is thereby not affected by the calibrations used during operation.
The summary data are also input to the simulations of the detector response.
In these simulations, 
the livetime of the detector 
as well as 
the status of the offshore data acquisition and 
count rate of each PMT are taken into account at any given time.
This reproduces
the operation of the detector and the conditions in the deep sea
according to the sampling of once per time slice which corresponds to a frequency of $10~\mathrm{Hz}$.

\section{Specifications}
\label{Specifications}

The data filter should sustain the data rates from the telescopes.
The maximal rate of data being written to disk is set to $25~\mathrm{Mb/s}$.
This corresponds to a reduction of the data by a factor of about $1 \times 10^{3}$.
The main event rate is due to muons produced by interactions of cosmic rays in the atmosphere above the detector but 
there is a finite probability of an accidental trigger
due to the background produced by decays of radioactive elements in the sea water and bioluminescence.
For optimal use of the data recording capacity, 
the rate of accidental triggers should be kept well below the expected event rate due to muons. 
The number of CPU cores in the servers required to process the data in real time is determined by the time to process a time slice.
To limit the costs and the size of the network, 
it is foreseen to operate each building block with less than $1000$ CPU cores with $15~\mathrm{HS06}$ per core.

%% file: Trigger.tex
\section{Data filter}
\label{Data filter}

The data from the neutrino telescope are processed by a set of data filters that run in parallel on the servers.
The data are collected first in the RAM of the CPUs of the servers. 
The collected data corresponding to a given time slice are processed as soon as the list of optical modules is complete.
When
the total amount of collected data or
the time difference between the earliest and latest data 
exceed predefined limits,
the data from the earliest time slice are processed anyway.
Possible intermittent delays in the network can thus be covered for as long as these can be accommodated in the available RAM.
The filtering of the data proceeds in several consecutive steps, commonly referred to as levels.

\subsection{Level-0}

By construction, the data from a PMT are sorted in time. 
Inside the FPGA, the data from the different PMTs inside the same optical module are multiplexed before they are sent to shore.
They are de\nobreakdashes-multiplexed at the start of the data processing on shore. 
While the data are de\nobreakdashes-multiplexed, the time stamp of each hit is corrected for the time calibration of the PMT.
For each optical module, 31 arrays of time sorted hits are obtained (one per PMT).
These hits are referred to as level\nobreakdashes-0 (L0) hits.

\subsection{Level-1}

In the level\nobreakdashes-1 (L1) step of the data processing,
coincidences between two (or more) L0 hits from different PMTs inside the same optical module are selected.
The 31 arrays with L0 hits are first multiplexed using a custom implementation of the merge sort algorithm.
An L1 hit is obtained when the time difference between two (or more) consecutive hits is less than a predefined time window.
A suitable value for the time window is $20~\mathrm{ns}$.
The time of the L1 hit is determined by the time of the earliest L0 hit.
Optionally, the time\nobreakdashes-over\nobreakdashes-threshold is included.
The time\nobreakdashes-over\nobreakdashes-threshold of the L1 hit is determined by
the first leading edge and
the last trailing edge of the L0 hits.
Multiple L1 hits within the same time window are combined to suppress double counting of hits.
The time of the first L1 hit is then maintained and the number of L0 hits incremented.
The typical rate of L1 hits per optical module amounts to about $1~\mathrm{kHz}$.
About half of this rate is due to genuine coincidences from decays of radioactive elements in the sea water and
a small fraction to the passage of muons \cite{km3net-PPMDU}.
The remainder can be attributed to random coincidences.

\subsection{Level-2}

In the level\nobreakdashes-2 (L2) step of the data processing,
the L1 data are subject to additional constraints.
The list of constraints includes
a time window,
a maximal space angle between the PMT axes and 
a minimal number of L0 hits.
To improve the signal\nobreakdashes-to\nobreakdashes-noise ratio, 
the time window is set to $10~\mathrm{ns}$,
the maximal space angle to $90~\mathrm{deg}$ and 
the minimal number of L0 hits to 2.
As a result, the rate of hits per optical module is reduced to about $700~\mathrm{Hz}$.

\subsection{Supernova}

Another selection of the L1 data is made for the detection of low\nobreakdashes-energy neutrinos ($\mathrm{MeV}$) from Supernovae that is parallel to the L2 step.
The time window and 
the maximal space angle between PMT axes are usually set to the same values as for the L2 selection
but
the minimal number of L0 hits is set to 5 instead.
As a result, the rate of hits per optical module is reduced to few $\mathrm{Hz}$.
A sizeable fraction of this rate is then due to muons produced in the atmosphere above the neutrino telescope \cite{km3net-PPMDU}.

\subsection{Summary}

For each time slice, a summary of the L0 data is made.
The rate of each PMT is computed by the number of hits divided by the duration of the time slice.
In case of a high\nobreakdashes-rate veto,
the time between the first and last hit is used instead.

\subsection{Trigger}

The trigger constitutes the final step in the data filtering.
Normally,
L2 hits are input to this step.
An event is triggered when a cluster of L2 hits is found which complies with predefined criteria.
For each event, all telescope data within a time window around the trigger are recorded.
This is referred to as the ``snapshot''.
The snapshot ranges from 
some extra time before the first hit to
some extra time after  the last  hit in the cluster.
To include all data that could be causally related to the event,
the extra time $\Delta T$ covers the largest diameter of the detector $D$.

\begin{eqnarray}
  \label{eq:Delta-T}
  \Delta T & = & \frac{n}{c} D
\end{eqnarray}

where
$c$ corresponds to the velocity of light in vacuum and
$n$ to the ratio thereof and the group velocity of light.
The latter depends on the wavelength of light and is around $1.35$.
For ORCA and ARCA, a suitable value for the extra time is $2.5~\mathrm{\mu s}$ and $10~\mathrm{\mu s}$, respectively.
The durations of the event and the snapshot are small compared to the duration of a time slice.
As a result, the impact of edge effects due the distribution of data is limited to few times $10^{-4}$.\\[\baselineskip]

The detection of an event proceeds in four main steps, namely:

\begin{enumerate}
\item sort hits in time;
\item for each primary hit, collect following hits within a predefined time window;
\item select hits that are causally related to the primary hit;
\item find largest subset of selected hits that are all causally related.
\end{enumerate}

The predefined time window in step 2 is the same as in equation \ref{eq:Delta-T}.
The general causality relation for a pair of hits can be formulated as follows:

\begin{eqnarray}
  \label{eq:causality}
  \left| t_i - t_j \right| & \le & \frac{n}{c} \left| \bar{x}_i - \bar{x}_j \right|  
\end{eqnarray}

where
$t_{i(j)}$ corresponds to the time of hit $i(j)$ and
$\bar{x}_{i(j)}$ to the position of the optical module.
The causality relation is not transitive.
As a consequence,
the guaranteed solution for finding the largest subset of hits that are all causally related 
from a set of $n$ hits requires $\mathcal{O}(n!)$ operations.
To limit the number of operations in step 4 to an acceptable level, 
a custom implementation of the clique algorithm is used \cite{CLIQUE}.
This algorithm requires $\mathcal{O}(n^2)$ operations.\\[\baselineskip]
A condition is applied to the number of hits and the number of optical modules which recorded the hits after steps 2--4.
When one of these numbers is less than the required minimum,
the processing of data continues with the next primary hit in step 2.
A suitable value for the minimal number of hits and optical modules to trigger an event is 5.
When the number of hits after step 2 is very large, 
the event is immediately triggered thereby avoiding waste of CPU time in steps 3--4.
This number is referred to as factory limit.
A suitable value is 100.\\[\baselineskip]

\subsection{Event signatures}

The rate of accidental triggers is determined by 
the number of optical modules involved and 
the probability that two hits accidentally comply with the causality relation.
When all optical modules in the detector are used, 
the combinatorial background is too large to reduce the data.
Therefore, 
two specific signatures are considered, namely
a muon traversing the detector and
a neutrino interacting inside the detector producing a shower of secondary particles without a muon.
These signatures are used in designated implementations of the trigger processing step.
The Cherenkov light from a muon can be described by a cone which propagates in a straight line with the speed of light in vacuum.
The Cherenkov light from a shower can be described by a sphere which expands with the group velocity of light in water.
The peak intensity of the light reduces between linearly (light cone) and quadratically (spherical) with the covered distance.
The light is further attenuated by absorption and scattering.
The range of Cherenkov light can therefore be limited without significant loss of the signal.
Because the range of a muon can be larger than that of light,
a longer distance between optical modules should be considered.
This distance can be decomposed into 
a part related to the muon and 
a part related to light \cite{BAKKER}.
The corresponding velocity is then $c$ and $c/n$, respectively.
This reduces the time window in equation \ref{eq:causality}.
The resulting volumes for a muon and shower correspond to a cylinder and a sphere, respectively.
For ARCA and ORCA, suitable values
for the radius of the cylinder are $120~\mathrm{m}$ and $45~\mathrm{m}$ and
for the diameter of the sphere $250~\mathrm{m}$ and $45~\mathrm{m}$, respectively.
For a given primary hit, the combinatorial background is then limited to the number of optical modules inside these volumes.\\[\baselineskip]

\begin{figure}[h!]
  \begin{center}
    \begin{picture}(16,7)
      \put(-0.5,-1){\includegraphics*[trim={0cm 0cm 0cm 0cm},clip,width=7.7cm]{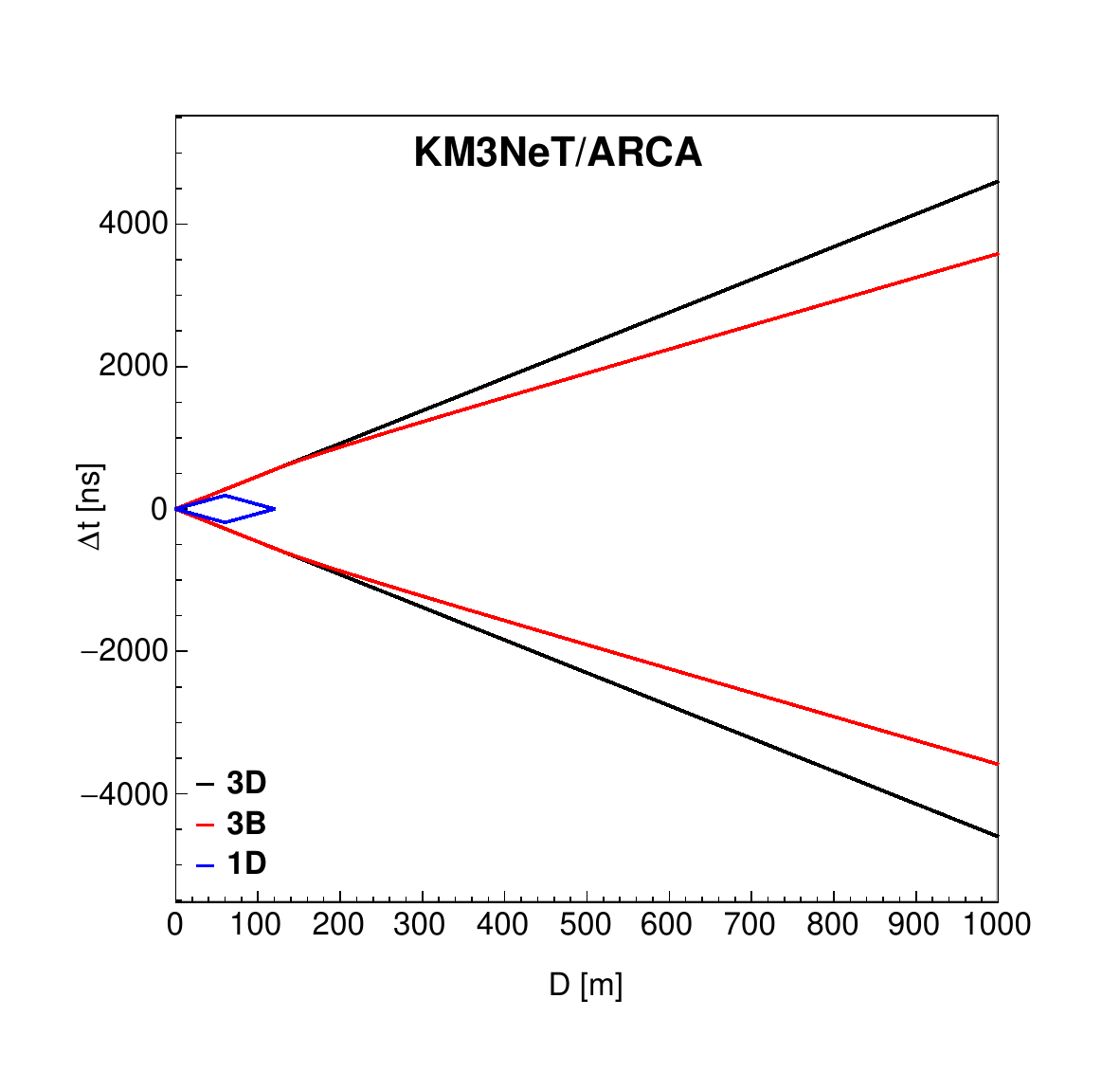}}
      \put(+6.6,-1){\includegraphics*[trim={0cm 0cm 0cm 0cm},clip,width=7.7cm]{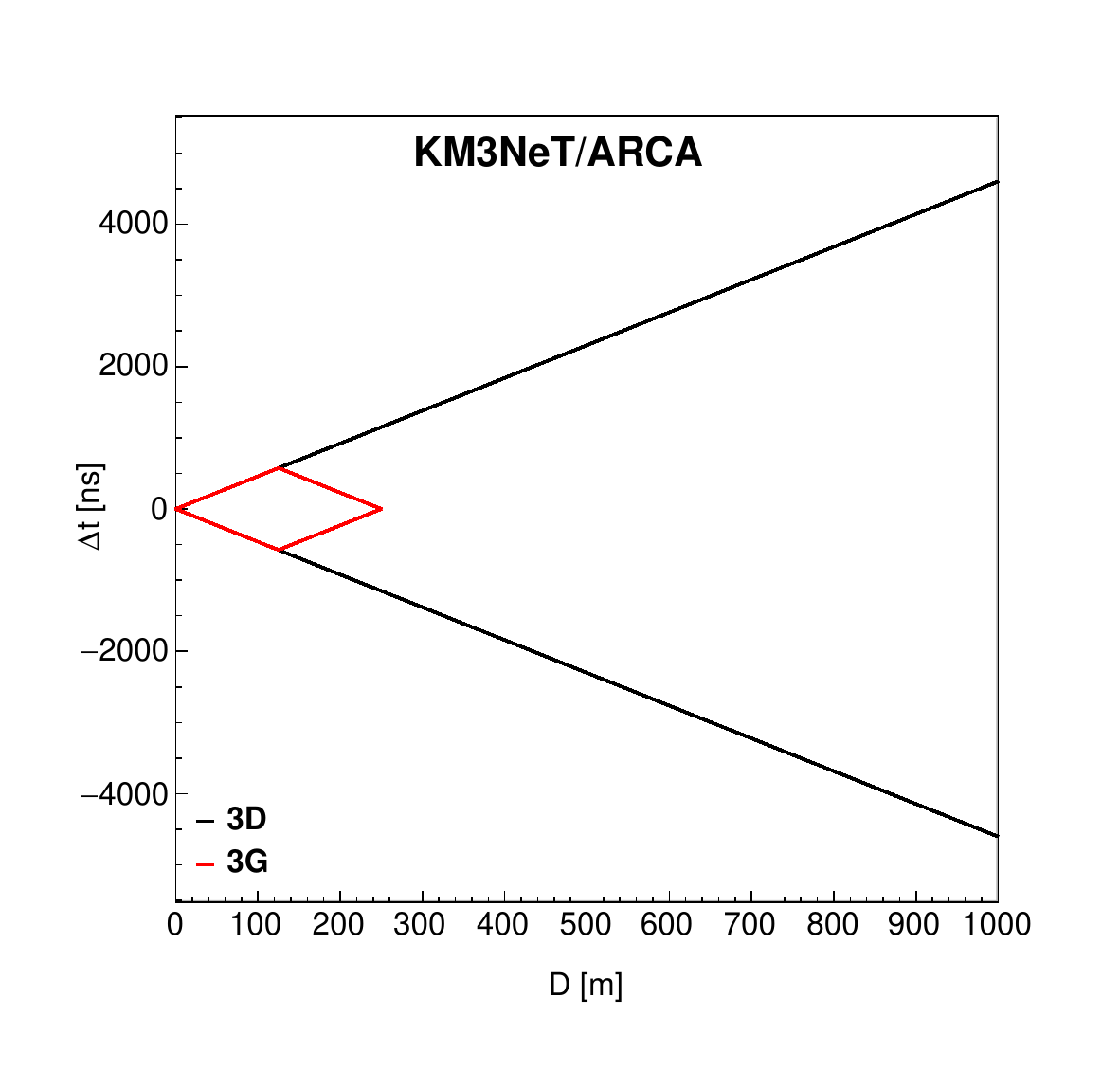}}
    \end{picture}
  \end{center}
  \caption{\label{fig:causality}
    The time windows for the KM3NeT/ARCA neutrino telescope as a function of the distance between optical modules for the different causality relations (see text).
    The left (right) figure applies to the signal from a muon (shower).
    The maximal distance roughly corresponds to the size of the detector.
    The label 3D corresponds to the general causality relation, 
    3B to the causality relation for which the distance covered by the signal is decomposed and
    1D to the causality relation for which the propagation time of the muon is taken into account.
    For the 1D case, the x\nobreakdashes-axis corresponds to the transverse distance with respect to the direction of the muon.
    The label 3G corresponds to the causality relation for which the distance covered by the signal is limited to the radius of a sphere.
  }
\end{figure}

By taking into account the propagation time of the muon, 
the time window is determined by the transverse distance between optical modules with respect to the muon direction.
The nearest\nobreakdashes-neighbour principle can thus be applied in two dimensions to optical modules that are far apart in the other dimension.
For a shower, 
the time window in equation \ref{eq:causality} increases up to the point that 
the distance between optical modules is equal to the assumed range of light. 
Beyond that point, 
the time window decreases because the signal should originate from somewhere in between the optical modules.
This counteracts the volumetric increase of the number of optical modules with the distance.
The reduced number of optical modules and time window lower the probability of an accidental coincidence by a factor of about $100$.
As a result, the rate of accidental triggers is reduced by a factor of $100^{m - 1}$ 
for a minimal number of hits to trigger an event equal to $m$.\\[\baselineskip]
The time windows for ARCA as a function of the distance between optical modules are shown in figure \ref{fig:causality} 
for the different causality criteria.
For ORCA, the same features as in figure \ref{fig:causality} apply but the distances are shorter and time windows smaller. 
Finally, a small amount of extra time is added to accommodate various uncertainties.
A suitable value for this extra time is $25~\mathrm{ns}$.\\[\baselineskip]

As a result of 
the application of a cylinder to the selection of optical modules and 
the restriction of the causality relation, 
the neutrino telescope is pointed to a position on the sky with a field\nobreakdashes-of\nobreakdashes-view of about $0.1~\mathrm{sr}$.
To increase the field\nobreakdashes-of\nobreakdashes-view, 
a grid of approximately equidistant directions is used.
With a spacing of $10~\mathrm{deg}$,
the full sky is covered using about 200 directions.\\[\baselineskip]

To lower the energy threshold for the detection of neutrino interactions in ORCA, 
an additional trigger is used which is based on the shower signature.
It requires a single L2 hit and a set of causally related L0 hits and is therefore referred to as mixed trigger.
To limit the rate of accidental triggers, 
an additional processing step is introduced after step 4.
In this step, 
a set of assumed vertex positions of the neutrino interactions is used.
These are distributed according a spherical grid around the L2 hit.
A suitable value for the radius of the sphere is $35~\mathrm{m}$ and for the spacing between the vertices $2~\mathrm{m}$.
For a given vertex position, 
the propagation time of the Cherenkov light from the vertex to the optical module can be taken into account.
The remaining distance in the causality relation is then limited to the spacing between the vertex positions.\\[\baselineskip]

The different implementations of the trigger processing step are applied to the same data.
A unique integer value is used to identify the trigger in use.
This value is included in the event data via a corresponding bit in a trigger mask.
When two (or more) events overlap in time, they are merged.
The lists of hits in the cluster and in the snapshot from the different events are then combined, avoiding duplications.
The resulting trigger mask is set to the logical OR of the underlying trigger masks.
The number of times that an event has been merged is also included in the event data.\\[\baselineskip]

\begin{table}[h!]
  \begin{center}
    \begin{tabular}{|l|l|r@{~}l|r@{~}l|}
      \hline
      level      &  parameter                &                 \multicolumn{4}{c|}{value}                  \\
                 &                           &  \multicolumn{2}{c}{ORCA}    &  \multicolumn{2}{c|}{ARCA}   \\
      \hline
      \hline
      L1         &  time window              &   $20$  &  $\mathrm{ns}$     &   $20$  &  $\mathrm{ns}$     \\
      \hline
      L2         &  time window              &   $10$  &  $\mathrm{ns}$     &   $10$  &  $\mathrm{ns}$     \\
      L2         &  angle between PMT axes   &   $90$  &  $\mathrm{deg}$    &   $90$  &  $\mathrm{deg}$    \\
      L2         &  number of L0 hits        &    $2$  &                    &    $2$  &                    \\
      \hline
      Supernova  &  time window              &   $10$  &  $\mathrm{ns}$     &   $10$  &  $\mathrm{ns}$     \\
      Supernova  &  angle between PMT axes   &   $90$  &  $\mathrm{deg}$    &   $90$  &  $\mathrm{deg}$    \\
      Supernova  &  number of L0 hits        &    $5$  &                    &    $5$  &                    \\
      \hline
      Trigger    &  extra time for snapshot  &  $2.5$  &  $\mathrm{\mu s}$  &   $10$  &  $\mathrm{\mu s}$  \\
      Trigger    &  number of L2 hits        &    $5$  &                    &    $5$  &                    \\
      ~~muon     &  radius of cylinder       &   $45$  &  $\mathrm{m}$      &  $120$  &  $\mathrm{m}$      \\
      ~~muon     &  spacing of directions    &   $10$  &  $\mathrm{deg}$    &   $10$  &  $\mathrm{deg}$    \\
      ~~shower   &  diameter of sphere       &   $45$  &  $\mathrm{m}$      &  $250$  &  $\mathrm{m}$      \\
      ~~mixed    &  number of L2 hits        &    $1$  &                    &  \multicolumn{2}{c|}{n.a.}   \\
      ~~mixed    &  radius for vertices      &   $35$  &  $\mathrm{m}$      &  \multicolumn{2}{c|}{n.a.}   \\
      ~~mixed    &  spacing of vertices      &    $2$  &  $\mathrm{m}$      &  \multicolumn{2}{c|}{n.a.}   \\
      \hline
    \end{tabular}
  \end{center}
  \caption{\label{tab:trigger}
    Main parameters of the data filter.
  }
\end{table}

The main parameters of the data filter are summarised in table \ref{tab:trigger}.
The quoted values are also suitable for the detectors that are currently in operation and can still be tuned.

%% file: Performance.tex
\section{Performance}
\label{Performance}

To evaluate the performance of the data filter, 
simulations of the detector response to interactions of neutrinos and the passage of muons are made \cite{km3net-gseagen}.
With these simulations, 
Monte Carlo data are generated that conform with the telescope data.
They are subsequently processed with the same trigger algorithms as the data from the neutrino telescopes.\\[\baselineskip]

\subsection{Purity}

\begin{figure}[h!]
  \begin{center}
    \begin{picture}(16,7)
      \put(-0.5,-1){\includegraphics*[trim={0cm 0cm 0cm 0cm},clip,width=7.8cm]{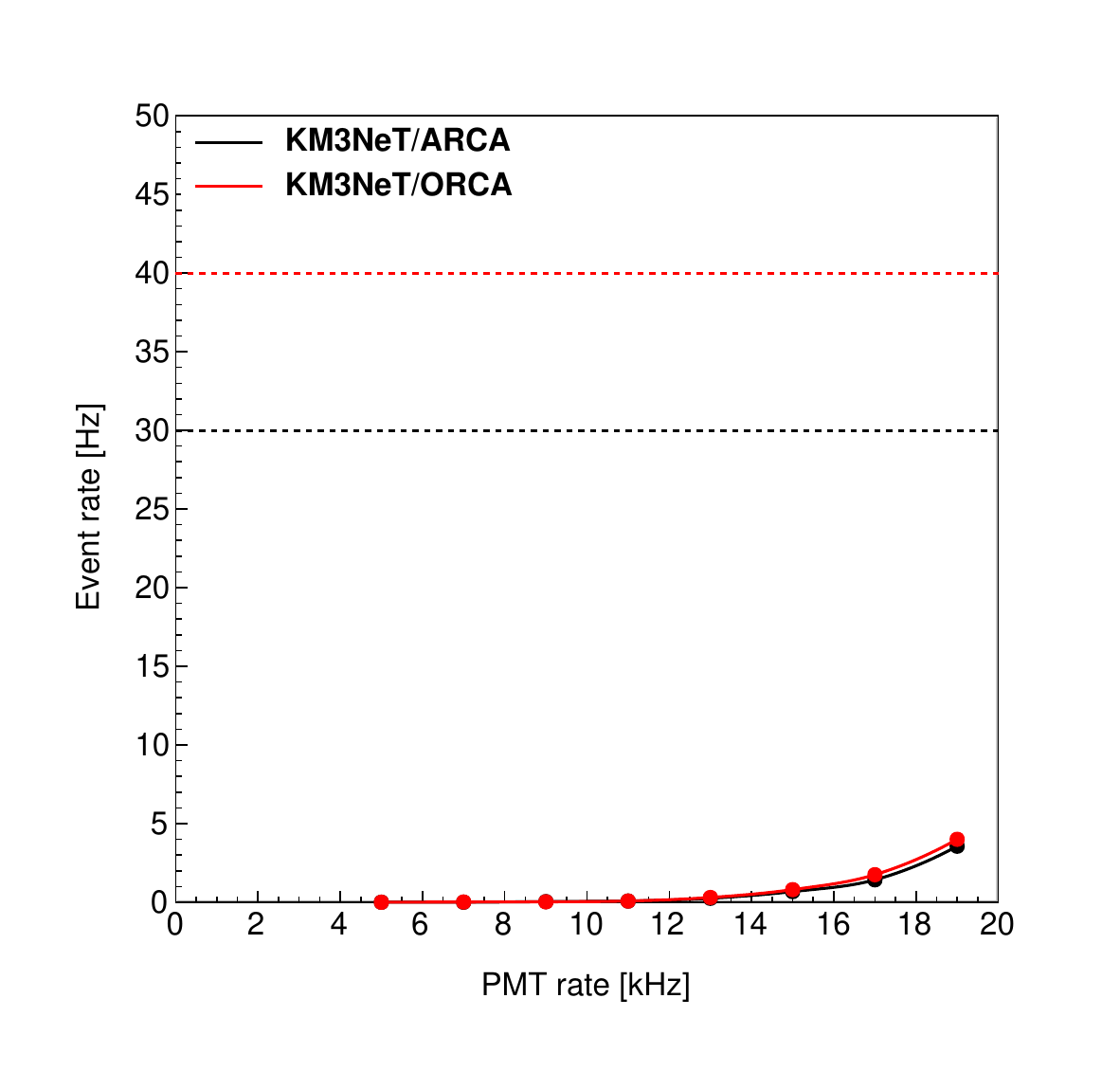}}
      \put(+6.5,-1){\includegraphics*[trim={0cm 0cm 0cm 0cm},clip,width=7.8cm]{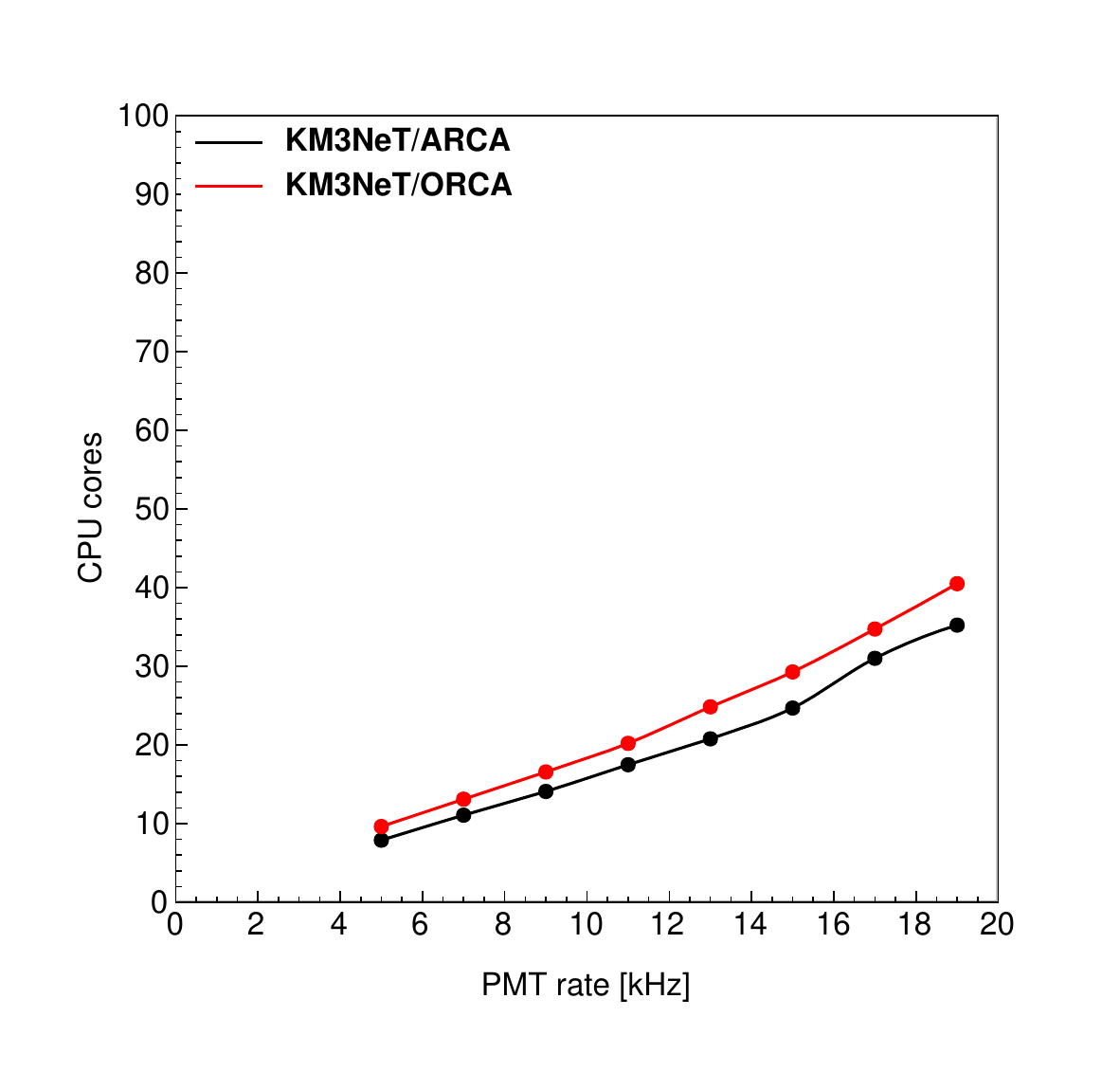}}
    \end{picture}
  \end{center}
  \caption{\label{fig:profile}
    The rate of events (left) and the required number of CPU cores (right) as a function of the singles rate of the PMTs for one building block of the KM3NeT/ARCA and KM3NeT/ORCA neutrino telescopes.
    The points correspond to the results obtained from a simulation of the random background.
    The dashed lines correspond to the expected rates of events due to muons produced by interactions of cosmic rays in the atmosphere above the detectors.
    A single CPU core of the \texttt{Intel(R) Xeon(R) E E-2478} processor is used.
  }
\end{figure}

The impurity of the data filter is defined as the ratio of 
the event rates caused by background and signal.
Here,
decays of radioactive elements in the sea water and bioluminescence are considered as the background and
muons produced by cosmic ray interactions in the atmosphere above the detector as the signal.
The signal is simulated by generating muons at the depth of the detector
according to parametrisations of the observed spectra \cite{MUPAGE}.
The expected event rate for one building block of ARCA and ORCA is found to be $30 \pm 5~\mathrm{Hz}$ and $40 \pm 5~\mathrm{Hz}$, respectively.
The quoted uncertainty covers the uncertainties about  
the incident flux,
the PMT characteristics and
the optical properties of the sea water.
The background is simulated by generating hits with random times.
The times of consecutive hits follow an exponential probability distribution according to a given rate.
The two\nobreakdashes-fold coincidence rate due to genuine coincidences
from decays of radioactive elements in the sea water is set to $500~\mathrm{Hz}$ and
the singles rate is varied from $5~\mathrm{kHz}$ to $20~\mathrm{kHz}$.
The upper limit corresponds to the high\nobreakdashes-rate veto that is implemented in the FPGA.
The simulations of the background are also used to determine the minimal number of CPU cores needed to sustain the rate of incoming data.
The rate of events and the required number of CPU cores are shown in figure \ref{fig:profile} as a function of the singles rate of the PMTs for one building block of ARCA and ORCA.\\[\baselineskip]

As can be seen from figure \ref{fig:profile},
the rate of events due to background is very small at the nominal rate of the PMTs.
The impurity of the data filter is then also very small ($< 1\%$).
At the maximal data taking rate of a PMT, the event rate is a few $\mathrm{Hz}$.
The corresponding impurity is then about $10\%$.

\subsection{Capacity}

The relative contributions to the CPU time of the processing steps in the data filter are summarised in table \ref{tab:cpu}.
The largest contribution is from the L1 processing step.
The contributions from the Calibration, L1 and L2 processing steps are common to the different trigger algorithms and depend linearly on the singles rate of the PMTs.
The trigger processing step involves the largest computational complexity but contributes only 20\% to the total CPU time.
The overhead and the contribution of signals to the CPU time are small.
As a result, the required number of CPU cores is less than $50$.
This complies with the specifications.\\[\baselineskip]

\begin{table}[h!]
  \begin{center}
    \begin{tabular}{|l|r|}
      \hline
      step                 &  CPU    \\
      \hline
      \hline
      Calibration          &   5\%   \\
      L1                   &  65\%   \\
      L2                   &   5\%   \\
      Trigger              &  20\%   \\
      overhead             &   5\%   \\
      \hline
    \end{tabular}
  \end{center}
  \caption{\label{tab:cpu}
    The relative contributions to the CPU time of the different processing steps in the data filter.
  }
\end{table}

The total rate of data being recorded is primarily determined by the event and summary data.
The recording of other data is normally configured in such a way that their contributions are negligible.
The data rate due to the summary data is proportional to the number of optical modules in the detector
because the sampling frequency and the data volume per optical module are fixed.
For one building block, it amounts to about $10~\mathrm{Mb/s}$.
The data rate due to the event data is determined by the product of the rate and data volume of the events. 
The latter is determined by the product of 
the duration of the snapshot, 
the total count rate of all PMTs in the detector and 
the size of a single hit.
The maximal data rate amounts to about $10~\mathrm{Mb/s}$ and $2.5~\mathrm{Mb/s}$ for ARCA and ORCA, respectively.
These values correspond to a worst\nobreakdashes-case\nobreakdashes-scenario.
The total data rate complies nonetheless with the specifications.\\[\baselineskip]

\subsection{Efficiency}

\begin{figure}[h!]
  \begin{center}
    \begin{picture}(16,13)
      \put(-0.5,+6){\includegraphics*[trim={0cm 0cm 0cm 0cm},clip,width=7.8cm]{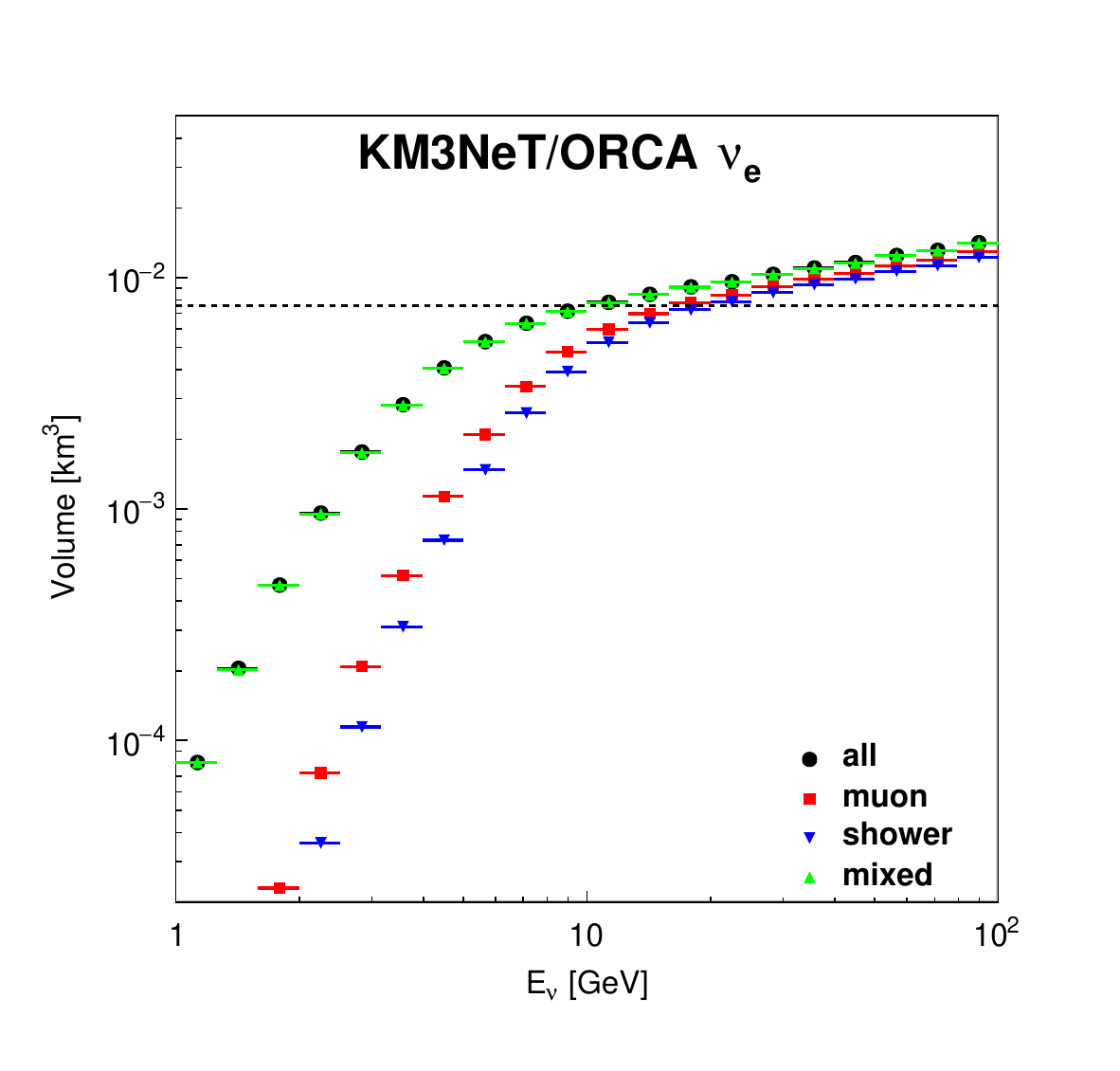}}
      \put(+6.6,+6){\includegraphics*[trim={0cm 0cm 0cm 0cm},clip,width=7.8cm]{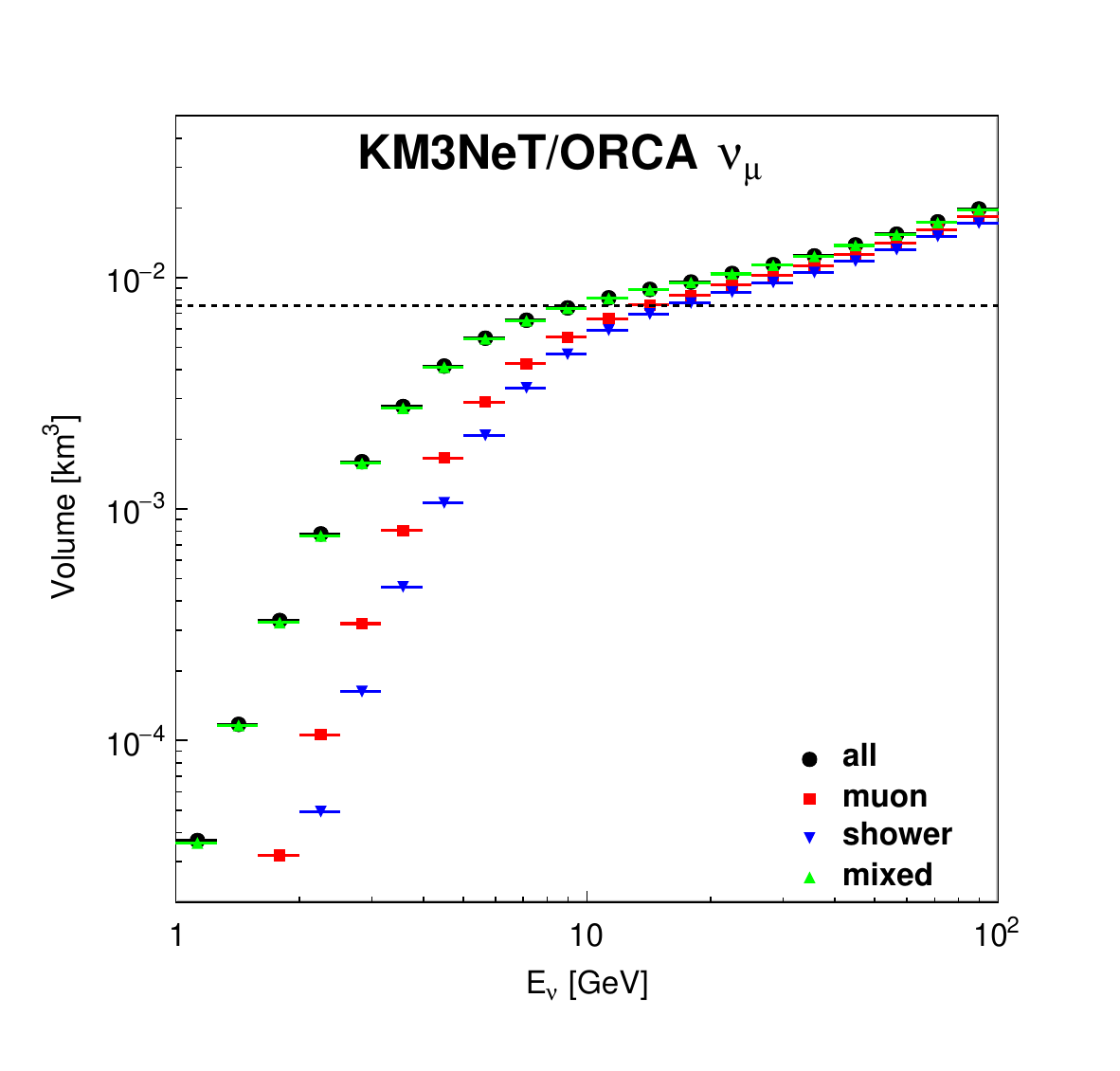}}
      \put(-0.5,-1){\includegraphics*[trim={0cm 0cm 0cm 0cm},clip,width=7.8cm]{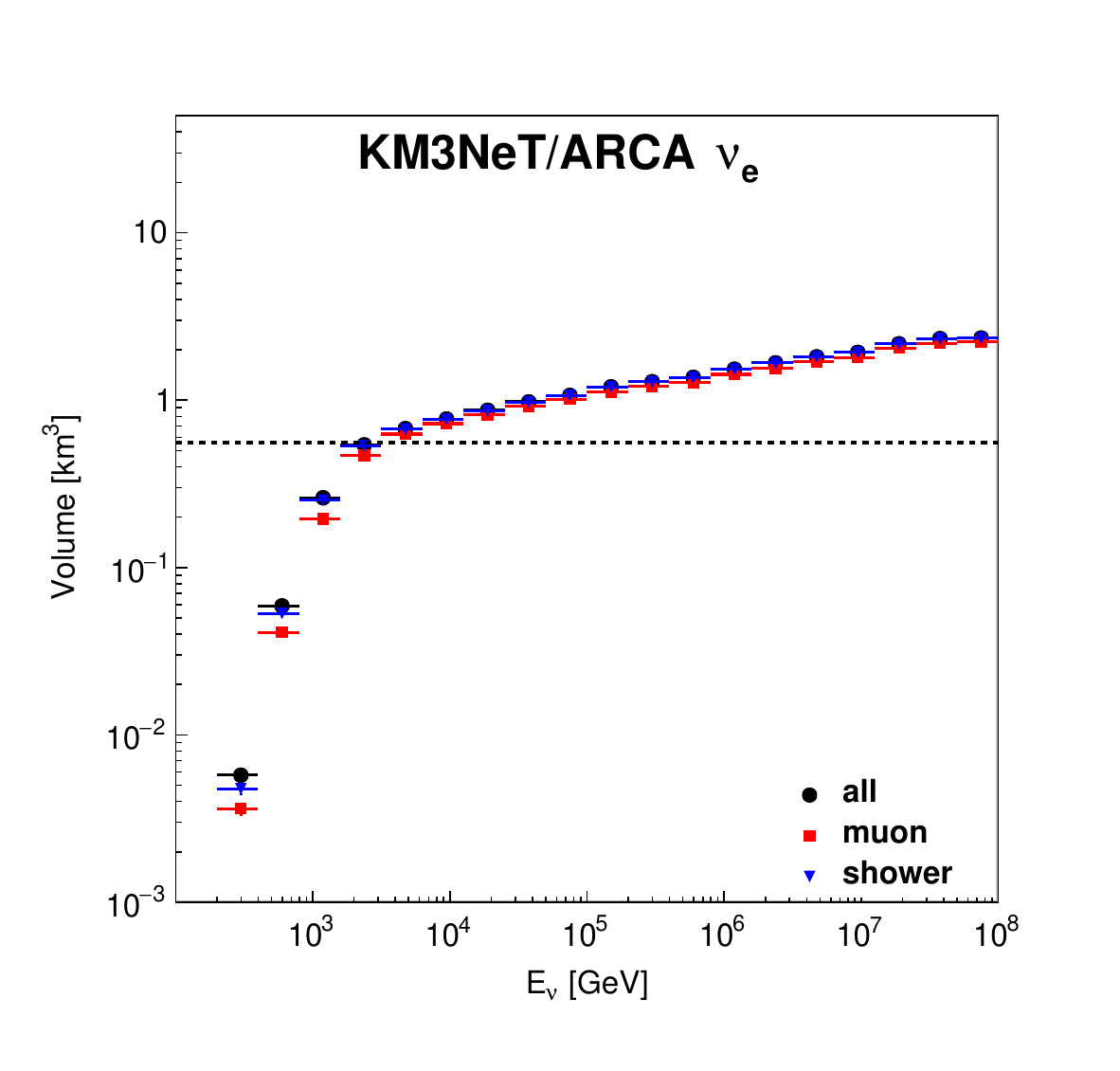}}
      \put(+6.6,-1){\includegraphics*[trim={0cm 0cm 0cm 0cm},clip,width=7.8cm]{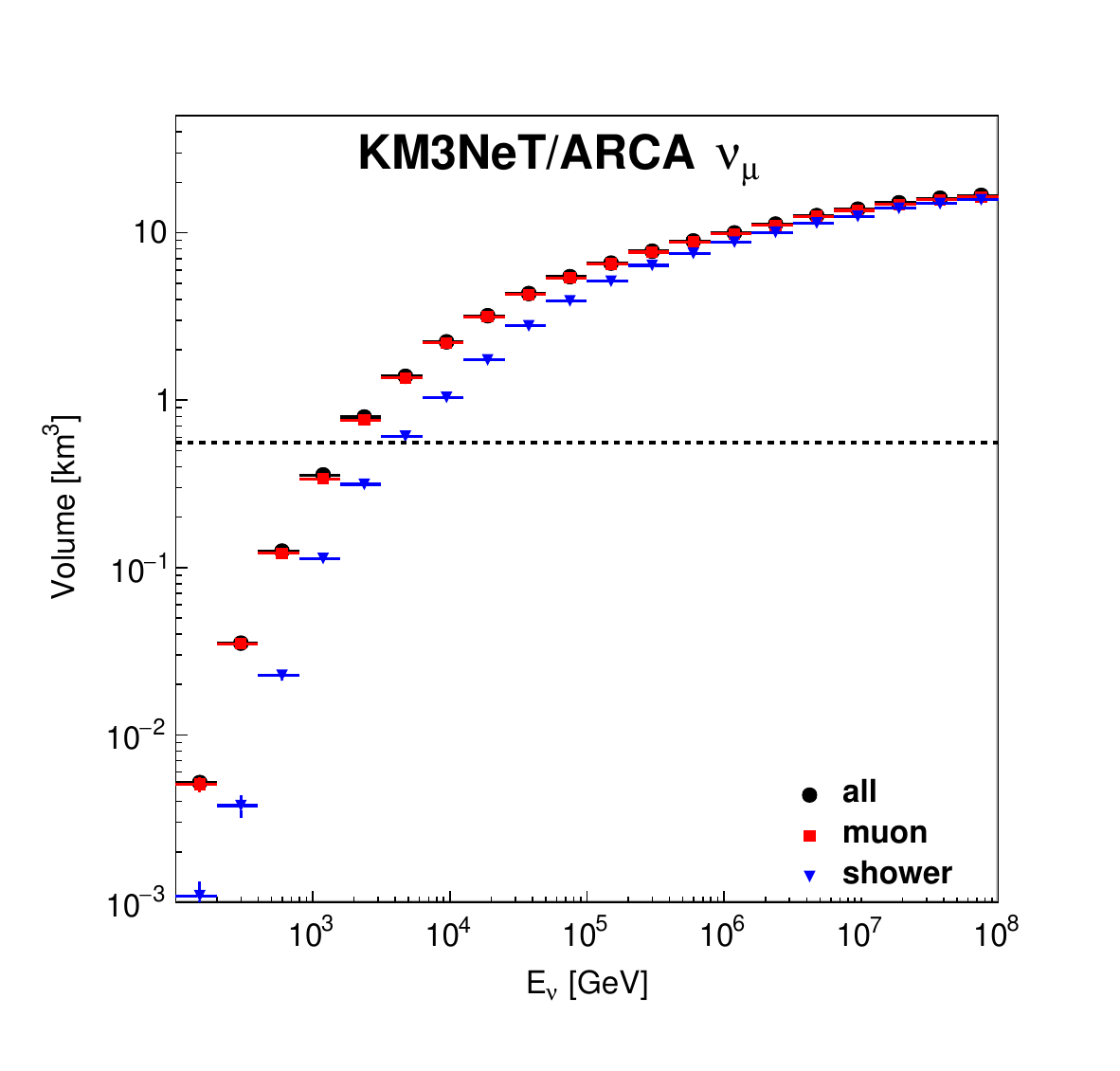}}
    \end{picture}
  \end{center}
  \caption{\label{fig:volume}
    The effective volume of one building block of the KM3NeT/ORCA (top) and KM3NeT/ARCA (bottom) neutrino telescopes 
    as a function of the neutrino energy and averaged over all directions 
    for charged\nobreakdashes-current interactions of electron (left) and muon (right) neutrinos (note the scale).
    The labels correspond to the signature that is used in the data filter.
    The dashed lines correspond to the geometrical volumes of the detectors.
  }
\end{figure}

The efficiency of the data filter can be quantified in terms of an effective volume. 
The effective volume is the volume in which the interaction of a neutrino will be triggered.
Each type of neutrino interaction yields a corresponding effective volume. 
As an example, 
the effective volumes of one building block of the ORCA and ARCA neutrino telescopes 
are shown in figure \ref{fig:volume} 
as a function of the neutrino energy 
for charged\nobreakdashes-current interactions of electron and muon neutrinos.
These volumes apply to a flux of neutrinos at the detector and have been averaged over all directions.
As a rule of thumb,
the charged\nobreakdashes-current interactions of anti\nobreakdashes-neutrinos yield a larger volume
because on average more energy is transferred from the incident neutrino to the outgoing charged lepton.
The neutral\nobreakdashes-current interactions of (anti\nobreakdashes-)neutrinos yield a smaller volume 
because on average less energy is transferred from the incident neutrino to particles producing Cherenkov light.
The charged\nobreakdashes-current interactions of tau (anti\nobreakdashes-)neutrinos yield a volume
that depends on the decay of the tau.
The corresponding effective volumes are similar to those of the muon or electron neutrino counterparts 
but generally smaller due to the invisible energy of the neutrinos produced in the decay of the tau.\\[\baselineskip]

As can be seen from figure \ref{fig:volume},
the effective volume reaches the geometrical volume at $10~\mathrm{GeV}$ and $1~\mathrm{TeV}$ for ORCA and ARCA, respectively.
The threshold can be attributed to the limited visible energy in the event.
The increase of the effective volumes beyond this point can be attributed to 
interactions of neutrinos in the vicinity of the detector.
The energy threshold of the ORCA detector is effectively lowered using the mixed trigger.\\[\baselineskip]
The figure\nobreakdashes-of\nobreakdashes-merit for the science objectives not only depends on the effective volume 
but also involves an offline analysis of the recorded data.
Because the criteria to detect events are tailored to the envisaged signals in the detector, 
the efficiency of the data filter comes with a positive bias in the selection of events in the offline analysis.

%% file: Conclusions.tex
\section{Conclusions}
\label{Conclusions}

A software data filter has been developed for the KM3NeT neutrino telescopes
which has been in use since the beginning of their operation, including the detection of KM3-230213A \cite{KM3-230213A}.
The interaction of a neutrino, 
the passage of a muon and 
other possible signals 
are detected when a cluster of hits is found in the data which complies with predefined criteria.
The background is suppressed by optimisation of the causality relations.
All telescope data within a time window around the detected events are recorded and archived for offline analyses.
In addition, summary, Supernova and calibration data are recorded.
Following an external alert, unfiltered or partially filtered data can be recorded which can recover up to several minutes of history.
The performance of the data filter has been evaluated in terms of the efficiency, purity and capacity.
The results of these evaluations comply with all specifications.
The design of the online data filter allows for 
tuning of the parameters and 
future implementations of alternative event triggers and data streams.
The all\nobreakdashes-data\nobreakdashes-to\nobreakdashes-shore concept makes it possible to maintain and upgrade
the computing infrastructure on shore during construction and operation of the KM3NeT neutrino telescopes.
The bandwidth of a standard fibre\nobreakdashes-optic network is sufficient to transfer the data.
The computing power of a modest farm of servers is sufficient to process the data.
As a result, a flexible solution exists to operate neutrino telescopes in the deep sea at low cost.

%% file: Acknowledgements.tex
\section{Acknowledgements}
The authors acknowledge the financial support of:
KM3NeT-INFRADEV2 project, funded by the European Union Horizon Europe Research and Innovation Programme under grant agreement No 101079679;
Funds for Scientific Research (FRS-FNRS), Francqui foundation, BAEF foundation.
Czech Science Foundation (GAČR 24-12702S);
Agence Nationale de la Recherche (contract ANR-15-CE31-0020), Centre National de la Recherche Scientifique (CNRS), Commission Europ\'eenne (FEDER fund and Marie Curie Program), LabEx UnivEarthS (ANR-10-LABX-0023 and ANR-18-IDEX-0001), Paris \^Ile-de-France Region, Normandy Region (Alpha, Blue-waves and Neptune), France,
The Provence-Alpes-Côte d'Azur Delegation for Research and Innovation (DRARI), the Provence-Alpes-Côte d'Azur region, the Bouches-du-Rhône Departmental Council, the Metropolis of Aix-Marseille Provence and the City of Marseille through the CPER 2021-2027 NEUMED project,
The CNRS Institut National de Physique Nucléaire et de Physique des Particules (IN2P3);
Shota Rustaveli National Science Foundation of Georgia (SRNSFG, FR-22-13708), Georgia;
This work is part of the MuSES project which has received funding from the European Research Council (ERC) under the European Union’s Horizon 2020 Research and Innovation Programme (grant agreement No 101142396).
The General Secretariat of Research and Innovation (GSRI), Greece;
Istituto Nazionale di Fisica Nucleare (INFN) and Ministero dell’Universit{\`a} e della Ricerca (MUR), through PRIN 2022 program (Grant PANTHEON 2022E2J4RK, Next Generation EU) and PON R\&I program (Avviso n. 424 del 28 febbraio 2018, Progetto PACK-PIR01 00021), Italy; IDMAR project Po-Fesr Sicilian Region az. 1.5.1; A. De Benedittis, W. Idrissi Ibnsalih, M. Bendahman, A. Nayerhoda, G. Papalashvili, I. C. Rea, A. Simonelli have been supported by the Italian Ministero dell'Universit{\`a} e della Ricerca (MUR), Progetto CIR01 00021 (Avviso n. 2595 del 24 dicembre 2019); KM3NeT4RR MUR Project National Recovery and Resilience Plan (NRRP), Mission 4 Component 2 Investment 3.1, Funded by the European Union – NextGenerationEU,CUP I57G21000040001, Concession Decree MUR No. n. Prot. 123 del 21/06/2022;
Ministry of Higher Education, Scientific Research and Innovation, Morocco, and the Arab Fund for Economic and Social Development, Kuwait;
Nederlandse organisatie voor Wetenschappelijk Onderzoek (NWO), the Netherlands;
The grant “AstroCeNT: Particle Astrophysics Science and Technology Centre”, carried out within the International Research Agendas programme of the Foundation for Polish Science financed by the European Union under the European Regional Development Fund; The program: “Excellence initiative-research university” for the AGH University in Krakow; The ARTIQ project: UMO-2021/01/2/ST6/00004 and ARTIQ/0004/2021;
Ministry of Research, Innovation and Digitalisation, Romania;
Slovak Research and Development Agency under Contract No. APVV-22-0413; Ministry of Education, Research, Development and Youth of the Slovak Republic;
MCIN for PID2021-124591NB-C41, -C42, -C43 and PDC2023-145913-I00 funded by MCIN/AEI/10.13039/501100011033 and
by “ERDF A way of making Europe”, for ASFAE/2022/014 and ASFAE/2022/023 with funding from
the EU NextGenerationEU (PRTR-C17.I01) and Generalitat Valenciana,
for Grant AST22\_6.2 with funding from Consejer\'{\i}a de Universidad,
Investigaci\'on e Innovaci\'on and Gobierno de Espa\~na and European Union
- NextGenerationEU, for CSIC-INFRA23013 and for CNS2023-144099,
Generalitat Valenciana for CIDEGENT/2020/049, CIDEGENT/2021/23, CIDEIG/2023/20, ESGENT2024/24, CIPROM/2023/51, GRISOLIAP/2021/192 and INNVA1/2024/110 (IVACE+i), Spain;
Khalifa University internal grants (ESIG-2023-008, RIG-2023-070 and RIG-2024-047), United Arab Emirates;
The European Union's Horizon 2020 Research and Innovation Programme (ChETEC-INFRA - Project no. 101008324).